\documentclass[twocolumn,
superscriptaddress,
prl,
showpacs,
preprintnumbers,
amsmath,amssymb]{revtex4}

\usepackage{graphicx,epsf,epsfig,ulem}% Include figure files
\usepackage{epsfig}
\usepackage{bm}% bold math
\usepackage{subfigure}
\usepackage{color}
\usepackage{graphicx}

\begin{document}

\title{Atomistic simulations of adiabatic coherent electron transport in triple donor systems}% Force line breaks with \\

\author{Rajib Rahman}
\affiliation{Network for Computational Nanotechnology, Purdue University, West Lafayette, IN 47907, USA}

\author{Seung H. Park}
\affiliation{Network for Computational Nanotechnology, Purdue University, West Lafayette, IN 47907, USA}

\author{Jared H. Cole}
\affiliation{Institute f\"ur Theoretische Festk\"orperphysik and DFG-Center for Functional Nanostructures (CFN), Universit\"at Karlsruhe, 76128 Karlsruhe, Germany}
\affiliation{Center for Quantum Computer Technology, School of Physics, University of Melbourne, VIC 3010, Australia}

\author{Andrew D. Greentree}
\affiliation{Center for Quantum Computer Technology, School of Physics, University of Melbourne, VIC 3010, Australia}

\author{Richard P. Muller}
\affiliation{Sandia National Laboratory, Albuquerque, NM 87185, USA}

\author{Gerhard Klimeck}
\affiliation{Network for Computational Nanotechnology, Purdue University, West Lafayette, IN 47907, USA} 
\affiliation{Jet Propulsion Laboratory, California Institute of Technology, Pasadena, CA 91109, USA}

\author{Lloyd C. L. Hollenberg}
\affiliation{Center for Quantum Computer Technology, School of Physics, University of Melbourne, VIC 3010, Australia}

\date{\today} 

\begin{abstract} 
A solid-state analogue of Stimulated Raman Adiabatic Passage can be implemented in a triple well solid-state system to coherently transport an electron across the wells with exponentially suppressed occupation in the central well at any point of time. Termed coherent tunneling adiabatic passage (CTAP), this method provides a robust way to transfer quantum information encoded in the electronic spin across a chain of quantum dots or donors. %By tuning the tunnel barriers between the wells by all-electrical means, CTAP effects adiabatic transfer of the particle between the ends of the chain. 
Using large scale atomistic tight-binding simulations involving over 3.5 million atoms, we verify the existence of a CTAP pathway in a realistic solid-state system: gated triple donors in silicon. Realistic gate profiles from commercial tools were combined with tight-binding methods to simulate gate control of the donor to donor tunnel barriers in the presence of cross-talk. As CTAP is an adiabatic protocol, it can be analyzed by solving the time independent problem at various stages of the pulse - justifying the use of time-independent tight-binding methods to this problem. %In addition to finding an adiabatic path for the triple donor system, the simulations were able to estimate typical adiabatic transfer times, and to verify that CTAP can occur despite the absence of the ideal localization assumption of the quantum optics treatment. 
Our results show that a three donor CTAP transfer, with inter-donor spacing of 15 nm can occur on timescales greater than 23 ps, well within experimentally accessible regimes. The method not only provides a tool to guide future CTAP experiments, but also illuminates the possibility of system engineering to enhance control and transfer times.     
\end{abstract} 

\pacs{05.60.Gg, 73.63.Kv, 73.21.La}

\maketitle{}
%Intro: Spin transport for QC
\section {I. Introduction}

The concept of encoding information in the superposition of quantum states offers revolutionary ways of performing computation and enormous improvement in speed and computing power for certain classes of algorithms {\cite{Nielsen.book.2000}}. Solid-state based quantum computer architectures have been the subject of much research due to their promise of scalability. Silicon systems are of particular interest because of the vast experience of the semiconductor industry in Si electronics and also because Si offers a relatively low noise environment for manipulating spins. There have been proposals to encode information in the nuclear {\cite{Kane.nature.1998}} or electronic spin {\cite{Vrijen.pra.2000}} of a phosphorus donor in Si, the orbital states of a singly ionized two-donor molecule {\cite{Hollenberg.prb.2004.1}}, the valley-split states of a Si quantum well or dot {\cite{Eriksson.NaturePhysics.2007}}, and in gate-confined 2DEGs {\cite{Loss.pra.1998}}. 

A potentially scalable quantum computer has to involve complex circuitry of qubits to perform multiple levels of error correction and fault-tolerance. In terms of the fault-tolerant threshold and defect tolerance such architectures will benefit from having separate zones for computation and measurement, and hence a mechanism for qubit transport {\cite{Hollenberg.prb.2006}}. Hence, the qubit state typically encoded in spin needs nonlocal transport while preserving the coherent superposition of the state vectors in the Hilbert space - the quantum mechanical equivalent of local bit transfers in traditional computers.   

%What is CTAP:
There already have been several proposals for non-local transport of encoded information in solid state quantum computers. Skinner \textit{et al.} {\cite{Skinner.prl.2003}} proposed a scheme in which electrons at an interface between Si and $\textrm{SiO}_{2}$ can be shuttled laterally along the surface by appropriate voltage pulses applied to a series of gates. This approach, however, requires a high gate density, and is susceptible to charge noise and spin-orbit interaction. Other approaches have used a chain of coupled harmonic oscillators {\cite{Plenio.prl.2004}} or interacting spins {\cite{Christandl.prl.2004, Friesen.prl.2007}} in the form of a quantum bus. 

In Ref {\cite{Greentree.prb.2004}}, a method was proposed to coherently transport quantum information encoded in the spin of an electron across a chain of quantum dots or ionized donors by modulating the tunnel barriers between them with voltage pulses. This technique is the solid-state analog of Stimulated Raman Adiabatic Passage (STIRAP) from quantum optics, and presents a robust population transfer mechanism by only electrical control. In this process, the electron is directly transferred from one end of the chain to the other with exponentially suppressed occupation in the middle of the chain at any point of time. This is made possible by adiabatically following certain pathways in the eigenspace connecting end states of the chain. This technique has been termed coherent tunneling adiabatic passage (CTAP) {\cite{Greentree.prb.2004}}. In addition to being a coherent transfer mechanism, this scheme does not change the energy of the electron, and is thus ideally a dissipationless technique. Moreover, this method requires gating only donors at the two ends of the chain, and hence can reduce the gate density of the architecture, although the three donor CTAP involves no advantages in gate density reduction. CTAP was incorporated in the bi-linear Si:P architecture design of Ref {\cite{Hollenberg.prb.2006}}.

Previous theoretical works on CTAP have investigated its feasibility for transporting single atoms {\cite{Eckert.pra.2004, Eckert.opt.2006}} and Bose-Einstein condensates {\cite{Graefe.pra.2006, Rab.pra.2008}}. CTAP has also been recently witnessed using photons in triple core optical waveguides {\cite{Longhi.pre.2006, Longhi.jphys.2007, Longhi.prb.2007}}. Recent papers {\cite{Cole.prb.2008, Tom.pra.2009}} made a thorough comparison between the quantum optics and solid-state versions of CTAP, highlighting the important differences between the two frameworks. %A simplistic model of 1D quantum wells was used to investigate some basic features of CTAP in the solid-state. 
Another recent work {\cite{Greentree.spie.2005}} showed that the time of electron transfer in CTAP scales as the square root of the number of dots in the chain, and the prospects for ion-implanted CTAP devices are discussed in Ref \cite{Donkelaar.arxiv.2009}.

%Significance of this work:
Here we demonstrate through numerical modeling the existence of CTAP in a realistic solid state system taking into account the atomistic nature of the underlying semiconductor. Our implicit goal is also to provide incentive and guidance to potential CTAP experiments in donors or quantum dots. However, we also provide results that show the utility of atomistic approaches to the solution of demanding time-domain quantum coherent problems. 

The test case investigated here involves a lattice of ${}^{28} \textrm{Si}$ atoms with three ionized donors and one bound electron under multiple gates, as shown in the schematic of Fig 1. This is a protoype case for CTAP in a long chain of donors as it involves most of the essential physics of the many-donor chain. 

CTAP in the quantum optics framework relies on an ideal localization assumption, and is well described by a three state system {\cite{Cole.prb.2008,Tom.pra.2009}}. In a solid state system under gate bias, this ideal localization assumption is no longer valid as the relevant states can admix with higher lying excited states. Since we diagonalize the full Hamiltonian of the system in an atomistic basis set, the excited states are explicitly included in our calculations, enabling us to verify whether CTAP can indeed occur in realistic solid-state systems once the ideal localization assumption is relaxed. 

Large scale atomistic tight-binding device simulations perfomed here also enable us to incorporate the Si host atoms in the model beyond the effective mass approximation and hence to include effects due to the full-bandstructure of the host. Furthermore, we utilize P donor models with valley-orbit interactions and core-corrections {\cite{Shaikh.encyclopedia.2008}}, and also use gate potentials obtained from commercial Poisson solvers to describe realistic devices. We are thus able to simulate gate control of tunneling barriers between the donors in the presence of gate crosstalk. The simulations carried out in this work are some of the most intensive single atom level quantum control simulations performed in a realistic solid-state system. 

Overall, the technique used here not only gives us a powerful simulation tool to model and guide future CTAP experiments, but also to show the existence of an adiabatic path for solid-state CTAP. This work also sets the stage for possible future investigations of the sensitivity of the adiabatic pathway to donor positioning and also of the scalability of the system with increasing number of donors. In Section II, we describe a typical CTAP device used in this work. Section III explains the concept of CTAP using a toy model. Section IV outlines the tight-binding method and relevant details. In Section V, we discuss the three-donor molecular spectrum and tunneling rates. Section VI describes pulse engineering to realize CTAP and reports adiabatic transfer times.      
 
  \begin{figure}[htbp]
  \centering
  \includegraphics[width=3.4in,height=3in]{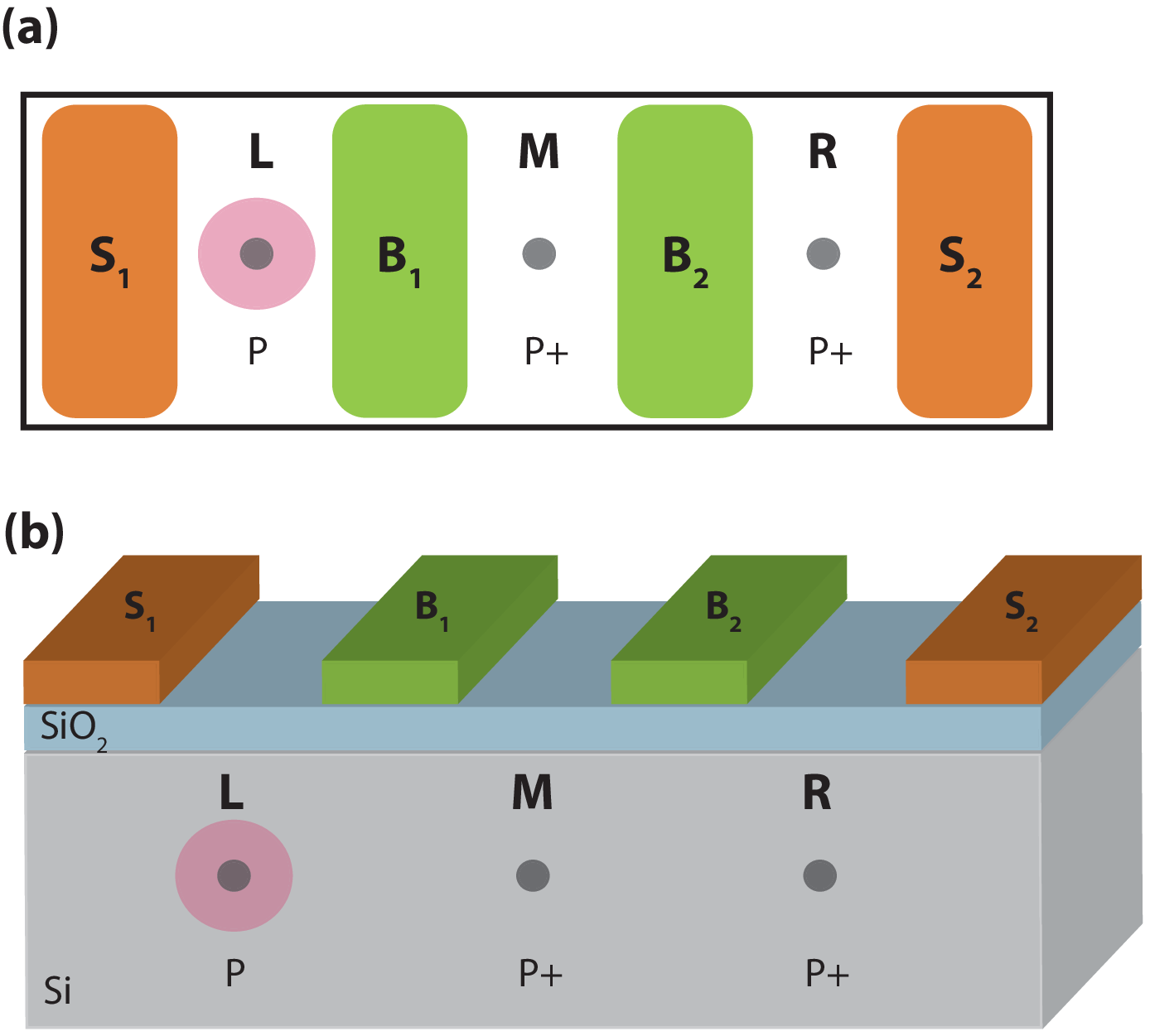}
  \caption{Schematic of the device for investigating CTAP. The barrier (B) gates modulate the tunneling barriers between the donors, while the symmetry (S) gates detune the left and right donor eigen states. The simulation has a 3D domain of 60.8 nm $\times$ 30.4 nm $\times$ 30.4 nm, and contains about 3.5 million atoms. The three donors are labeled L (left), M (middle), and R (right). a) Top view b) 3D view.}
  \label{ch5:1}
\end{figure}
  
% \begin{figure}[htbp]
 %\centering
 % \includegraphics[width=3.4in,height=3in]{CTAP_fig1.pdf}
 % \caption{Schematic of the device for investigating CTAP. The barrier (B) gates modulate the tunneling barriers between the donors, while the symmetry (S) gates detune the left and right donor eigenstates. The simulation has a 3D domain of 60.8 nm x 30.4 nm x 30.4 nm, and contains about 3.5 million atoms. The three donors are labeled L (left), M (middle), and R (right). a) Top view b) 3D view.}
%\end{figure}

%Device structure:
\section {II. Device structure}
%\subsection {Device Structure}
%\subsubsection {Device Structure}

A typical CTAP device design used in this work is shown in Fig. 1. The quantum mechanical eigenstates are computed in a cubic domain of 60.8 nm $\times$ 30.4 nm $\times$ 30.4 nm with comprising of about 3.5 million Si atoms with one electron bound across three ionized P donors. The modeling of this device uses an atomistic grid of 3D geometry. 

Two symmetry gates and two barrier gates are placed above a 5 nm thick oxide layer on top of the Si lattice. The function of the barrier (B) gates is to modulate the tunnel barriers between the donors. The symmetry (S) gates are used to detune the energies of the end donors, adding an extra degree of control for the CTAP pathway. The gates considered here are of 10 nm width. %as larger domains increase compute times without adding anything new to the analysis. 
Due to the relative closeness of the gates in this test device, there is significant cross-talk between them, making control relatively hard for this small device. Gate cross-talk in this context means that a typical gate can affect parts of the device in addition to its intended function. For example, the barrier gate $\textrm{B}_1$ is intended to control the tunnel barrier between the left (L) and the middle (M) donor only, but in reality it also affects both the barrier between the middle and the right (R) donor and the detuned energies of the end donors, L and R, relative to M. In a larger device, the gates will be farther apart, reducing the cross-talk effects and easing controllability. 

The donors are buried 15 nm below the oxide, and are also placed 15 nm apart from each other in the [100] direction. Relatively small device dimensions were chosen as a large number of tight-binding simulations had to be performed for different voltage configurations to zone in on the adiabatic path. However, the dimensions used here are sufficient to capture all the essential physics of the donor chain. Increasing the device domains further only leads to an increase in compute times without adding new effects to the analysis. The choice of 15 nm distance between a donor and an interface is due to the fact that a donor located more than 15 nm away from any interface in this model can be considered as an isolated donor in bulk Si free from interface effects {\cite{Rahman.iedm.2008}}. In contrast, an ideal CTAP device may have the impurities buried deeper and farther apart, and would involve a chain of donors in place of the middle donor. 

\section {III. Effective 3 $\times$ 3 model of CTAP} 

%A toy model for CTAP (3 x 3 CTAP model, Intutive and counter intuitive pulsing, adiabatic manipulation at low temperature):

The concept of CTAP is best described by a simple effective model. Assuming three different donor sites, and a wavefunction localized in each donor, we can use a 3 $\times$ 3 Hamiltonian describing the system in this 3-state basis. The Hamiltonian $H$ is of the form,
 %The diagonal terms of this matrix represent on-site energies of the donors, while off-diagonal terms represent coupling between two donors. 

%\begin{displaymath}
\begin{equation} \label{eq:H1} 
H =
\left( \begin{array}{ccc}
E_L & t_{LM} & t_{LR} \\
t_{LM}* & E_M & t_{MR} \\
t_{LR}* & t_{MR}* & E_R
\end{array} \right)
\end{equation}
%\end{displaymath}

\noindent
where $E_i$ is the on-site energy of $i$-th impurity, and $t_{ij}$ is the tunneling matrix element from impurity $i$ to impurity $j$. We can further simplify the system by assuming the ground state of the donors are aligned in energy so that $E_L = E_M = E_R $, and arbitrarily set the eigenvalues to 0. %$E_1 = 0$, and $E_2 - E_1 = \Delta$. 
We can also assume only nearest donor coupling by setting $t_{LR} = 0$. The reduced matrix is of the form,

%\begin{displaymath}
\begin{equation} \label{eq:H2} 
H =
\left( \begin{array}{ccc}
0 & t_{LM} & 0 \\
t_{LM}* & 0 & t_{MR} \\
0 & t_{MR}* & 0
\end{array} \right)
\end{equation}
%\end{displaymath}

%\begin{equation} \label{eq:hamiltonian} 									
%\end{equation}

The eigenvalues and eigenvectors of this matrix are of the form, 
\begin{align}								
&E_1 = -\sqrt{| t_{LM} |^2 +| t_{MR} |^2} \\ \nonumber \\
&E_2 = 0 \\ \nonumber \\
&E_3 =  +\sqrt{| t_{LM} |^2 +| t_{MR} |^2}
\end{align}

\begin{align}								
&|\Psi_1 \rangle = \frac{t_{LM} |L\rangle -\sqrt{| t_{LM} |^2 + | t_{MR} |^2} | M \rangle +t_{MR} | R \rangle}{\sqrt{2 \left( | t_{LM} |^2 +| t_{MR} |^2 \right)}} \\ \nonumber \\
&|\Psi_2 \rangle= \frac{t_{MR} |L\rangle - t_{LM} |R \rangle}{\sqrt{| t_{LM} |^2 +| t_{MR} |^2} } \\  \nonumber \\
&|\Psi_3 \rangle= \frac{t_{LM} |L\rangle +\sqrt{| t_{LM} |^2 + | t_{MR} |^2} | M \rangle+ t_{MR} | R \rangle}{\sqrt{2 \left( | t_{LM} |^2 +| t_{MR} |^2 \right) } } 
\end{align}

We initialize the system by localizing the electron around the left donor (setting $t_{LM}=0$ to prevent hybridization). Looking at Eqs. 6-8, we see that this configuration corresponds to the second dressed state $|\Psi_2 \rangle$. In a real system, setting $t_{LM} = 0$ involves raising the tunneling barrier between L and M by applying a more negative bias to $\textrm{B}_1$, while adjusting other gates to compensate for cross-talk. In practice, this also means that $|t_{LM}| \ll |t_{MR}|$ so that the barrier between M and R is lower than that between L and M. At some point later we set the tunnel barriers $t_{LM}=t_{MR}$, at which point $|\Psi_2 \rangle$ represents a superposition of the L and R localized states (with no appreciable population in state M). Finally, we reduce $t_{MR}$ smoothly to zero leaving the system localized in state R. Following such a sequence allows us to adiabatically evolve the system from a L localized state to a R localized state, without populating the central donor. This adiabatic pathway is described in the ideal limit by simply following state $|\Psi_2 \rangle$ from initial to final state. During the entire transfer process, $E_2$ is held fixed at 0, resulting in no change of electron energy and no acquisition of any dynamical phase. Since this is an adiabatic transfer process at a very low temperature, the electron always occupies the state in which it started.

The pulsing sequence described above is quite counter-intuitive in nature. If we are trying to transfer the electron from L to R, then an intuitive pulsing sequence would involve lowering the barrier between L and M first, and then the barrier between M and R. This will transport the electron first to the middle well, and then to the right well, very much like a bucket-brigade device. However, if the barriers are modulated in the reverse order such that the barrier between M and R is lowered first and then that between L and M according to the CTAP protocol, then the electron is transferred directly from L to R in a much more robust fashion \cite{Greentree.prb.2004} in terms of pulse control over tunnel rates. A signature of the CTAP protocol is an exponentially suppressed occupation at the middle donor.

%However, if the barriers are modulated in the reverse order such that the barrier between M and R is modulated first and then that between L and M according to the CTAP protocol, then the electron is transferred directly from L to R with exponentially suppressed occupation at the middle donor. %This method is more robust as the electron does not interfere with the charge noise of the donor chain.   

\section {IV. Method} 

The tight binding model employed in this work is the 20 band $sp^{3}d^{5}s*$ spin model with nearest-neighbour interactions. This model incorporates spin inherently in the basis by duplicating the 10 spatial orbitals per atom for up and down spins. Spin-orbit interactions of the host are also included by onsite $p$-orbital spin-orbit corrections {\cite{Chadi.prb.1977}}. The model parameters were optimized by a genetic algorithm with appropriate constraints to reproduce the important features of the bulk bandstructure of the host {\cite{Slater.physrev.1954, Boykin.prb.2004, Klimeck.cmes.2002}}. The P donors were modeled by Coulomb potentials screened by the dielectric constant of Si. At the donor site, a cut-off potential $\textrm{U}_0$, was used, and its value optimized so that the ground state binding energy of -45.6 meV was obtained for a donor in bulk Si. In this model, the magnitude of $\textrm{U}_0$ reflects the strength of the valley-orbit interaction responsible for lifting the six-fold degeneracy of the $1s$ manifold of the impurity. It was shown in an earlier work {\cite{Shaikh.encyclopedia.2008}} that the splitting between the singlet, triplet and doublet components of the $1s$ manifold increase with the magnitude of $\textrm{U}_0$. This semi-emprical highly optimized technique was able to reproduce the full single donor spectrum very accurately. 

The electrostatic gate potentials were obtained from a commercial Poisson solver, and was then interpolated into the atomistic grid for the tight-binding simulations. Due to the large number of computer-intensive simulations required to home in on the adiabatic path, it was not possible to generate each time the total potential profile of all the gates taken together. Instead, we generated the potential profiles of each gate separately, and assumed the net potential can be obtained from the superposition principle. Some non-linear behavior is expected to arise from the additional fringing fields near the gates when multiple gates are turned on at the same time. However, such non-linear behavior is expected only to add small voltage corrections. Furthermore, the basic principle of tuning the tunnel barriers between the donors by barrier gates to realize the adiabatic pathway remains unchanged.   

Closed boundary conditions with a model of dangling bond passivation was used to model the interfaces. The full Hamiltonian of about 3.5 million atoms including the four gate potentials was solved by parallel Lanczos and Block Lanczos algorithm to capture the relevant eigenvalues and wave functions. Typical computation time for 6 states was 7 hours on 40 processors {\cite{nanohub.note}}. Although CTAP is a time dependent problem requiring transient voltage pulses, it can be analyzed by snapshots of the wave-functions at different biases obtained from the time independent Schr\"odinger equation.  

The tight-binding method under the hood of the Nano-electronic Modeling Tool (NEMO-3D) {\cite{Klimeck.cmes.2002, Klimeck.ted.2007}} was used previously to verify the Stark shift coefficients of the hyperfine interaction of the donor spin {\cite{Rahman.prl.2007}} with respect to ESR measurements {\cite{Bradbury.prl.2006}}. The method was also used to compute orbital Stark shifts of an As donor in Si close to the oxide barrier, and could explain energy level measurements from transport experiments in commercial FINFETs {\cite{Rogge.NaturePhysics.2008}}. The same method was successfully applied to investigate valley-splitting with alloy-disorder and step roughness in Si quantum wells {\cite{Neerav.apl.2007}}, and was also used to model quantum dots for optical communication wavelengths {\cite{Usman.nano.2008}}.

 \begin{figure}[htbp]
  \centering
  \includegraphics[width=3.4in,height=2.6in]{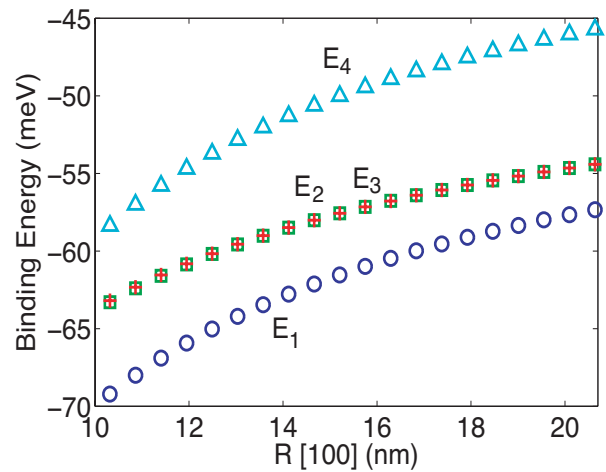}
  \caption{The energies of the 4 lowest single electron states of the 3P$^{2+}$ donor molecule as a function of donor separation in [100]. The energies are shown relative to the conduction band minima. The energy differences between the states correspond to tunnel couplings between the donors.}
  \label{ch5:2}
\end{figure}

%\begin{figure}[htbp]
%  \centering
 % \includegraphics[width=3.4in,height=2.6in]{CTAP_Fig2.png}
 % \caption{The energies of the 4 lowest single electron states of the 3P$^{2+}$ donor molecule as a function of donor separation in [100]. The energies are shown relative to the conduction band minima. The energy differences between the states correspond to tunnel couplings between the donors.}
%\end{figure}

\section {V. Three-donor molecule at zero gate bias} 

A single phosphorus donor in bulk Si has a ground state (GS) of $\textrm{A}_1$ symmetry at -45.6 meV relative to the conduction band (CB) minima. Above this, there is an orbital triplet manifold of $\textrm{T}_2$ symmetry at -33.9 meV, and an orbital doublet manifold of $\textrm{E}_1$ symmetry at -32.6 meV {\cite{Ramdas.progphys.1981}}. When three ionized donors are located close-by, coupling between the wells produce molecular states that may span over the whole chain. Fig. 2 shows the energies of the four lowest states of the donor molecule (3P$^{2+}$) at zero gate bias as a function of separation distance along [100]. The energy differences between these states are proportional to the tunnel barriers. The separation distance between the donors is incremented in equal steps so that the L and R are always equidistant from M. 

It has been shown that the tunnel coupling for a two-donor (2P$^{+}$) charge qubit exhibits oscillatory behaviour with relative donor separations along [110] and [111] {\cite{Hu.prb.2005}}. The exchange coupling between the donor electrons of a two-donor molecule (2P) has also been shown to exhibit oscillations with donor positions, a consequence of phase pinning of the Bloch functions at the donor sites {\cite{Koiller.prl.2001}}. This has posed some controllability issues for two-qubit operations of the Si:P based qubits \cite{Kane.nature.1998, Vrijen.pra.2000, Hollenberg.prb.2004.1}, although individual qubit characterization is likely to resolve the problem. It would be interesting to investigate if the CTAP protocol is also susceptible to radial and angular donor misalignments, but this goes well beyond the scope of our present work. 

 \begin{figure}[htbp]
  \centering
  \includegraphics[width=3.4in,height=3.8in]{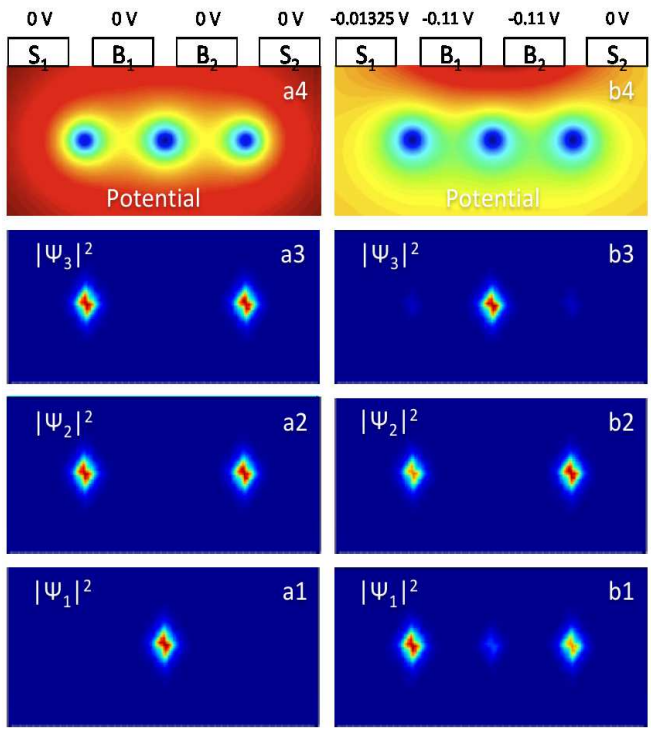}
  \caption{Particular potential landscapes (a4 and b4) and the corresponding molecular states (a1-3 and b1-3) of the triple-donor device. The left column shows the donor molecule at zero gate bias. The right column shows the donor molecule at the middle of the adiabatic path.  Note the almost zero electron density at the middle donor in the 2nd eigenstate.}
  \label{ch5:3}
\end{figure}

%\begin{figure}[htbp]
 % \centering
%  \includegraphics[width=3.4in,height=3.8in]{Fig3.png}
 % \caption{Potential landscape (top row) and Molecular states (bottom 3 rows) of the triple-donor device. The left column shows the donor molecule at zero gate bias. The right column shows the donor molecule at the middle of the adiabatic path.  Note the zero electron density at the middle donor in the 2nd eigenstate.}
%\end{figure}

In a triple Coulomb well generated from the superposition of three isolated Coulomb potential, the middle well is deeper than the left and right wells. As a result, the ground state of the system will have electron density at the middle donor. The L and R wells are essentially at the same energy, and form nearly degenerate bonding and anti-bonding states. Top row of Fig 3 shows the total device potential including the donors and the gates, while the remaining rows show the three lowest wavefunctions of 3P$^{2+}$. The left column of Fig. 3 portrays the donor molecule at zero gate bias. The wavefunctions conform to the symmetries described above. At low gate bias, these lowest three states arise from linear combinations the single $\textrm{A}_1$ states from each donor. Similarly, the closely spaced orbital triplet and doublet manifolds from each impurity will interact to form 15 excited molecular states. Since there is a gap of about 11 meV between the orbital singlet and the orbital triplet manifold of a single impurity, there will also be a modest energy gap between the lowest three molecular states and the higher states. However, this gap is likely to decrease as donor separations decrease, and we need to make sure that the manifold of the lowest three states is sufficiently isolated from the higher states for ideal CTAP operation. Fig. 2 also shows that the centrally occupied state $\textrm{E}_1$ approaches $\textrm{E}_2$ and $\textrm{E}_3$ as donor separation increases and the system moves towards the isolated donor regime, in which all the three $\textrm{A}_1$ states are degenerate. $\textrm{E}_4$ is the lowest molecular state arising from the $\textrm{T}_2$ manifold. Even if the gate lengths can be scaled down to nanometers, donor separations less than 10 nm are not desirable as the $\textrm{E}_4$ state approaches the lowest three manifold. However, donor separations of 15 nm or more seem reasonable for CTAP.

\section {VI. Gate pulsing to realize CTAP} 

To obtain the wavefunction symmetries for CTAP described in Section III, the molecular states needed to be aligned close to each other in energy. This was achieved by applying negative biases to the barrier gates so that the middle donor is pulled up in energy close to $\textrm{E}_2$ and $\textrm{E}_3$. This also lowers the effective tunnel barriers between the donors causing more hybridization between the states. In the right column of Fig 3, we applied -0.11 V to each of the barrier gates, and compensated for gate crosstalk by applying a small bias to the symmetry gate $\textrm{S}_1$. The wavefunctions indeed correspond to the symmetries identified from the effective 3 $\times$ 3 model. However there is only a limited range of barrier gate biases where this happens. In the simulations, we noticed a barrier gate bias window between -0.1 V and -0.12 V where the states are strongly interacting to produce the molecular symmetries we seek. 

\begin{figure}[htbp]
  \centering
  \includegraphics[width=3.6in,height=4.2in]{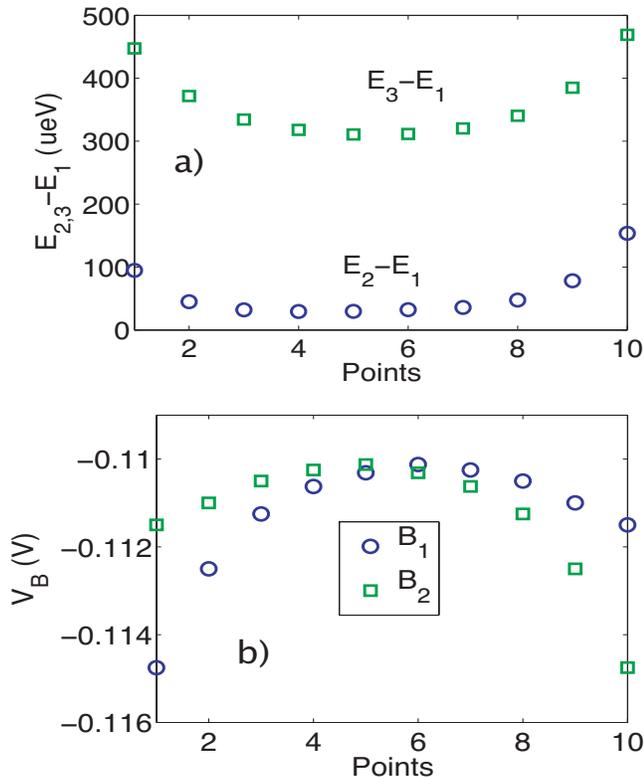}
  \caption{(a) Evolution of the lowest three eigenvalues with the CTAP gate pulse. The gap between $E_1$ and $E_2$ is minimum at point 5. (b) The $\textrm{B}_1$ and $\textrm{B}_2$ gate voltages at the 10 points of the adiabatic path. At all points, the $\textrm{S}_1$ and $\textrm{S}_2$ gates are held fixed at -0.01325 V and 0 V respectively to compensate for cross-talk.}
  \label{ch5:4}
\end{figure}

%\begin{figure}[htbp]
  %\centering
  %\includegraphics[width=3.6in,height=4.2in]{CTAP_fig5.pdf}
  %\caption{(a) Evolution of the lowest three eigenvalues with the CTAP gate pulse. The gap between $E_1$ and $E_2$ is minimum at point 5. (b) The $\textrm{B}_1$ and $\textrm{B}_2$ gate voltages at the 10 points of the adiabatic path. At all points, the $\textrm{S}_1$ and $\textrm{S}_2$ gates are held fixed at -0.01325 V and 0 V respectively to compensate for cross-talk.}
%\end{figure}

\begin{figure}[htbp]
  \centering
  \includegraphics[width=3in,height=5in]{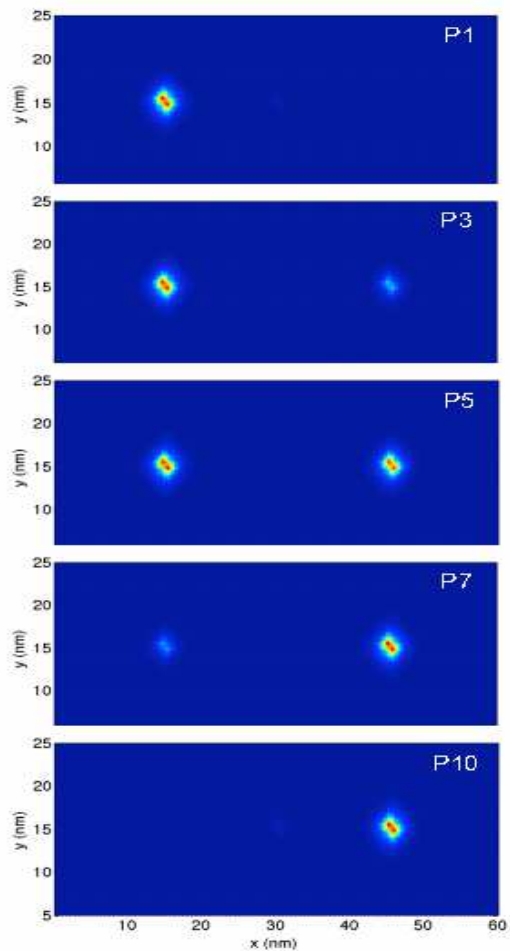}
  \caption{The 1st excited state under a voltage sequence that transfers the electron from the left to right donor without middle occupation. The voltage configurations involve modulations of $\textrm{B}_1$ and $\textrm{B}_2$ gate voltages, while keeping $\textrm{S}_1$ and $\textrm{S}_2$ fixed. Only 5 points of the adiabatic path are shown here. For the barrier gate voltages at these points, refer to Fig 4(b).}
  \label{ch5:5}
\end{figure}

%\begin{figure}[htbp]
%  \centering
%  \includegraphics[width=3in,height=5in]{Fig4_CTAP.png}
 % \caption{The 1st excited state under a voltage sequence that transfers the electron from the left to right donor with minimal middle occupation. The voltage configurations involve modulations of $\textrm{B}_1$ and $\textrm{B}_2$ gate voltages, while keeping $\textrm{S}_1$ and $\textrm{S}_2$ fixed. Only 5 points of the adiabatic path are shown here.}
%\end{figure}

With small barrier gate modulation around -0.11 V, we were able to obtain other points on the adiabatic trajectory. In Fig 4, we show the eigenvalues (4a) and the barrier $\textrm{B}_1$ and $\textrm{B}_2$ gate voltages (4b) at 10 points of the adiabatic path. We observe anti-crossing between the states $\textrm{E}_1$ and $\textrm{E}_2$ at the middle of the path in Fig 4a. The energy gap between the lowest two states approaches a non-zero minimum of 29 $\mu$eV at the middle of the path at point 5. From this minimum gap, we can roughly estimate the electron transfer timescale as $\tau=\frac{\hbar}{\Delta_{min}}=0.023$ ns for a distance of 30.4 nm. %This transfer time can also be decreased further by pushing $\textrm{E}_2$ closer to $\textrm{E}_3$, which can occur at less negative barrier gate voltages.  
This minimum energy gap depends both on barrier gate voltages and donor separations, and can be engineered to achieve faster transfer times.

Fig 5 shows the electron localization corresponding to the first excited state at various stages of the electron transfer. The population gradually diminishes in the left donor and reappears in the right, with minimal leakage to the middle donor. This verifies for the first time that CTAP can be observed in a realistic solid-state system such as an interacting donor chain. Unlike the solution of the effective 3 $\times$ 3 model outlined in Section III, the more realistic simulations show that the CTAP state has a non-zero and time-varying energy. However, this is still essentially the same CTAP protocol described by the effective 3 $\times$ 3 model. 

%Fig 5a shows the evolution of the eigenvalues at various points in the adiabatic path.          
%Gate pulsing (eigenspectrum showing adiabatic path, time of electron trasnfer and distance of transfer):
%Translation between TB and 3 x 3 CTAP model (makes future work easier):
%The adiabatic pathway in this work was found by a combination of educated guesswork and trial and error, and required a large number of time-consuimng quantum simulations. In future works involving straggle and potential experiment modelling, it will be convienent to establish a procedure that can help to zone in on the adiabatic path faster and reduce the number of large scale simulations. This can be done by mapping the tight binding results on to the 3 x 3 model shown earlier. If we use the same simulation domain in tight-binding, but solve for each impurity at a time at zero gate bias, we can obtain three single donor A1 states. We can use these three donors as basis  
%Spin orbit

\section {VII. Conclusion} 

We have demonstrated the possibility of CTAP in a triple donor chain through precise numerical modeling. Originally developed in the quantum optics framework, CTAP relies on an ideal localization assumption and a few state based description, which are generally not met in solid-state systems. We have shown that a realistic solid-state system can still exhibit CTAP when the few state description is abandoned and a Schroedinger wave description is used with many molecular states considered in the calculations. Despite controllability problems due to gate crosstalk in a small device and band-structure effects of the host material, it is possible to find adiabatic trajectories that define CTAP. The large scale highly optimized quantum mechanical device simulations done here not only show the existence of adiabatic pathways in a triple donor system, but also helps to devise a technique to model and guide potential experiments. The results enable us to estimate typical adiabatic transfer timescale of 0.023 ns for this device with left and right impurities separated by 30.4 nm. Since CTAP presents a robust and coherent method to transport electronic spin in a quantum circuitry, experimental demonstration of CTAP in solid-state systems should be sought after. 

Although the three-donor case serves well as a test, the real benefit of CTAP will be evident in a long donor chain. Under suitable gate pulses, the donor electron can be transported from one end of the chain to another, carrying along with it the quantum information encoded in its spin. Realizing such a system will indeed help to solve some of the critical information transport problems in solid state quantum computing architectures. It is therefore necessary to investigate scalability of the adiabatic pulsing scheme to increasing number of donors. Further studies need to be undertaken to investigate the sensitivity of the adiabatic path to relative donor positioning and also to investigate spin-orbit coupling effects at various stages of the transfer. 

\begin{acknowledgments}
This work was supported by the Australian Research Council, the Australian Government, and the US National Security Agency (NSA), and the Army Research Office (ARO) under contract number W911NF-04-1-0290. Part of the development of NEMO-3D was initially performed at JPL, Caltech under a contract with NASA. NCN/nanohub.org computational resources were used in this work. Sandia is a multiprogram laboratory operated by Sandia Corporation, a Lockheed Martin Company, for the United States Department of Energy's National Nuclear Security Administration under contract DE-AC04-94AL85000. JHC wishes to acknowledge the support of the Alexander von Humboldt Foundation. ADG and LCLH acknowldege the Australian Research Council for financial support (Projects No. DP0880466 and No. DP0770715, respectively). We also acknowledge comments from Dr. Malcolm Carroll of SNL.
\end{acknowledgments}
Electronic address: rrahman@purdue.edu

%\vspace{-0.5cm}


\begin{thebibliography}{100}
\bibitem{nanohub.note} nanoHUB.org computational resource of a 256-node 3.3GHz Pentium Irvindale PC cluster was used in this work. The tight-binding calculations were done using the Nano-Electronic Modeling Tool (NEMO-3D).
\end{thebibliography}
\end{document}